# Visually Wired NFTs: Exploring the Role of Inspiration in Non-Fungible Tokens

**Lucio La Cava, Davide Costa, Andrea Tagarelli**

Dept. Computer Engineering, Modeling, Electronics, and Systems Engineering (DIMES)
University of Calabria, 87036 Rende (CS), Italy
{lucio.lacava, davide.costa, tagarelli}@dimes.unical.it

## Abstract

The fervor for Non-Fungible Tokens (NFTs) attracted countless creators, leading to a Big Bang of digital assets driven by latent or explicit forms of inspiration, as in many creative processes. This work exploits Vision Transformers and graph-based modeling to delve into visual inspiration phenomena between NFTs over the years. Our goals include unveiling the main structural traits that shape visual inspiration networks, exploring the interrelation between visual inspiration and asset performances, investigating crypto influence on inspiration processes, and explaining the inspiration relationships among NFTs. Our findings unveil how the pervasiveness of inspiration led to a temporary saturation of the visual feature space, the impact of the dichotomy between inspiring and inspired NFTs on their financial performance, and an intrinsic self-regulatory mechanism between markets and inspiration waves. Our work can serve as a starting point for gaining a broader view of the evolution of Web3.

Non-Fungible Tokens (NFTs) have gained remarkable popularity in recent years as a type of digital asset ensuring uniqueness and traceability through the so-called *smart-contracts*, i.e., code-based automatic and self-executing contracts. Due to their pervasiveness, NFTs have become the leading application of blockchain technology and pioneers of the *Web3*'s advent. Thanks to their ability of self-regulating and tracing the history of each asset, from its *minting* (i.e., creation) to all its trades, NFTs have the potential to revolutionize the way we trade any asset that can be digitally represented. Such power has been suddenly recognized by a plethora of digital creators and corporations, which embraced NFTs to authenticate assets from different sources, such as images, video, and audio.

Investors and traders enthusiastically welcomed the expansion of this blockchain-based technology, as testified by a trading volume that exceeded $2 billion USD already in the first quarter of 2021.[1] Notable examples leading to unprecedented investments in the art world[2] include the sale of the first Tweet for more than $2.9 Million,[3] automatically generated pixeled artworks known as *CryptoPunks* selling for up to $24 Million,[4] or cartooned cats jamming the Ethereum network due to disproportionate sales.[5] Also, the sale of the first natively digital NFT-based work of art (Beeple's Everydays: The First 5000 Days) by Christie's auction house achieved the third highest price of $69.3 Million.[6]

The effects of such a disruptive technology are already tangible or under study in several domains (Gonserkewitz, Karger, and Jagals 2022), such as art and collectibles, game development, healthcare, preservation of cultural heritage.[7] Furthermore, NFTs represent the best-in-case technology today for the development of the *Metaverse* (Wang 2022; Wang et al. 2023; Ritterbusch and Teichmann 2023), acting as deeds of ownership for virtual lands or as forerunners of digital fashion (Joy et al. 2022).

While leading to countless investments, the NFT technology has also attracted the attention of the research community, which resulted in a body of studies that have been proposed in the past few months (Wang et al. 2021; Nadini et al. 2021; Mekacher et al. 2022; Kapoor et al. 2022; von Wachter et al. 2022; Meyns and Dalipi 2022). Most existing studies have focused on three main topics: NFT price prediction or forecast (Nadini et al. 2021; Mekacher et al. 2022; Ho et al. 2022; Dowling 2022; Pinto-Gutiérrez et al. 2022; Ante 2022; Apostu et al. 2022; Anselmi and Petrella 2023; Costa, La Cava, and Tagarelli 2023), crypto influence on the NFT markets (Kapoor et al. 2022), and NFT buyer-seller relations (Vasan, Janosov, and Barabási 2022; Colavizza 2022).

**Contributions.** Nonetheless, several questions still remain unanswered about the visual nature of NFTs and their implications on the market. None of the aforementioned works has, in particular, investigated the underlying patterns that can be induced from the raw visual features of the NFTs — especially across the boundary of collections owned by different creators — and how such patterns can have effect on the creation of new NFTs, and hence on the NFT market. By contrast, our proposed study aims to fill a gap in the literature by addressing the above aspects.

[1] https://nonfungible.com/reports/2021/en/q1-quarterly-nft-market-report

[2] https://decrypt.co/62898/most-expensive-nfts-ever-sold

[3] https://www.cnbc.com/2021/03/22/jack-dorsey-sells-his-first-tweet-ever-as-an-nft-for-over-2point9-million.html

[4] https://decrypt.co/92819/cryptopunks-ethereum-nft-sells-for-nearly-24m-doubling-previous-record

[5] https://qz.com/1145833/cryptokitties-is-causing-ethereum-network-congestion/

[6] https://www.christies.com/about-us/press-archive/details?PressReleaseID=9970

[7] https://cointelegraph.com/news/the-world-s-cultural-heritage-is-being-preserved-one-nft-at-a-time

A major goal of our work is to leverage visual features learned from NFTs in order to build a suitable network model for capturing latent visual influences that some NFTs can exert on others. In particular, we investigate a special type of visual influence that can be detected *whenever an NFT appears to be visually close to another that was published earlier in the market*. We call this phenomenon **visual inspiration**, which is conceptualized to be as general enough to also include potential extreme cases, such as copying or even plagiarism. To the best of our knowledge, we are the first to deal with this challenging topic.

Throughout this paper, we shall investigate the visual inspiration mechanisms underlying the NFT landscape to date, through answering the following main research questions:

- **(RQ1)** *Caught in the net* — Can we effectively model NFT images and their pairwise similarity relations to unveil possible patterns of NFT visual inspiration? (Sect. *Data Extraction and Network Modeling*)
- **(RQ2)** *Just the way they are* — What are the main structural traits exhibited by the inferred NFT inspiration networks? (Sect. *Analysis of the NFT and Collection Networks*)
- **(RQ3)** *To be or not to be... inspired?* — Are inspired and inspiring NFTs performing differently on the market? (Sect. *Market-based Characterization of the NFT Visual Inspiration Phenomenon*)
- **(RQ4)** *Crypto-Flu* — What is the correlation between crypto markets and inspiration processes in the NFT landscape? (Sect. *Crypto Influence Dynamics*)
- **(RQ5)** *Tell me why!* — What are the most relevant visual features to explain the inspiring relation between two NFTs? (Sect. *Explainability Aspects of the NFT Visual Learning Model*)

**Scope and Limitations.** The data used in this study are largely representative in terms of diversification and coverage of the NFT landscape. In fact, they contain NFTs from heterogeneous platforms — thus with different scopes — and cover the market history up to the steady-state of mid-2021, enabling us to capture the main consolidated fluctuations that appeared in the NFT selling scenario. Nonetheless, being out of the temporal range of our data, extreme events that appeared in the NFT market in 2022 are not considered in this study.

We also point out that our data contain timestamped information on the NFTs' selling prices, however they lack the assets' minting times, i.e., release times. We do not treat this missing point as a major inconvenience, since our focus is on those NFTs that have actually drawn attention in the market rather than those published but that might not be noticed by traders.

Also, we are aware of the risk that some visual inspiration processes might be altered or even "boosted" from multiple (fake) accounts belonging to a single person; however, delving into such phenomena would shift the focus to users, while the scope of this work involves visual features of NFTs. This issue is worthy of attention and we consider it part of our future work.

## Data Extraction and Network Modeling

This section describes the data used in our study and our defined NFT and graph representation models.

**Data.** We use the largest and most representative dataset concerning the NFT landscape to date (Nadini et al. 2021). It contains 6.1M purchase transactions involving 4.7M NFTs and more than 4k collections observed between 2017 and 2021. NFTs are grouped by the authors into six main categories, which indicate their scope, namely Art (18.46%), Collectible (28.85%), Games (47.21%), Metaverse (0.1%), Utility (0.17%), and Other (5.21%) — values within parenthesis are percent proportions. Each transaction in the dataset is associated with metadata that allows keeping track of the actors involved in the transaction and the main information associated with each NFT, such as its selling price, collection and category, and image URLs. After filtering out inaccessible NFTs or NFTs with less than one sell, our final dataset consisted in more than 180k NFT images of various formats (.png, .jpeg and .webp), and corresponding metadata.

**NFT visual representation model.** To represent NFT images, we produce dense vectorial representations (embeddings), whose feature space is learned via *Transformer-based pre-trained visual models* (PVMs). These models, which allow avoiding manual or domain-driven selection of prominent features, are indeed the "de-facto" standards in computer vision. More specifically, we exploit a PVM based methodology for NFT visual feature extraction that has shown to be effective in (Costa, La Cava, and Tagarelli 2023); note that, however, the purpose of exploitation of visual feature extraction in this work is different from the one in (Costa, La Cava, and Tagarelli 2023), which focused on a task of NFT financial performance prediction.

Let us denote with $\mathcal{I}$ the set of NFT images available in our dataset. Any NFT image $n_i \in \mathcal{I}$ can conveniently be represented as a *token sequence* $\mathcal{T}_i = [\tau_{i,0}, \tau_{i,1}, ..., \tau_{i,|\mathcal{T}_i|}]$, where $\tau_{i,j}$ denotes the $j$-th token of image $n_i$. Given a pre-trained visual model $\mathsf{PVM}$, image tokens are densely encoded into a $d$-dimensional latent space, that is, $\mathsf{PVM}(\mathcal{T}_i) \in \mathcal{R}^{|\mathcal{T}_i| \times d}$. The resulting token embeddings are then fed through a pooling function $pooling(\cdot)$ in charge of producing, for each NFT image $n_i$, a single $d$-dimensional embedding $\mathbf{h}_i = pooling(\mathsf{PVM}(\mathcal{T}_i))$. A common approach, widely used in state-of-the-art PVMs is to get the output embedding of the special [CLS] token, whose pooling is generally considered a comprehensive representation of the entire (tokenized) input image.

Our choice for $\mathsf{PVM}$ in this study falls on *Vision Transformer* (ViT) (Dosovitskiy et al. 2021; Caron et al. 2021). ViT is the first vision representation learner capable of achieving very competitive results against state-of-the-art CNN based architectures (e.g., Inception-V3). Being a BERT-like architecture, ViT works by tokenizing an image into patches of a given size (e.g., $16 \times 16$ pixels, as in our used implementation)[8] so that a token sequence can be attended to by the *self-attention* mechanism, which is able to draw relationships between patches. ViT models have shown to excel in image classification tasks reaching up to 90% accuracy on ImageNet; in particular, our used implementation of ViT was pre-trained on ImageNet-21k (Ridnik et al. 2021).

It should be noted that our design choice for PVM is to use it as is, without any adaptation to the NFT domain; in other terms, our PVM was used in inference on NFTs, but not further trained on NFTs. Although this may appear as a limitation, we instead defined such a setting for the following reason: a visual model that has never seen digital assets, like NFTs, is more likely to mimic the newbie user who is approaching the NFT market, and as such s/he does not have domain-specific knowledge.

**Graph representation models for NFT visual inspiration.** Once computed the NFT image embeddings $\mathbf{h}$, we use them to model proximity graphs, both at NFT level and collection level, which will lead us to answer our **RQ1**; in the following, we provide the corresponding definitions.

NFT GRAPH. We define the *NFT visual-inspiration graph* as a directed weighted graph $\mathcal{G} = \langle V, E, w, ts, T_s, T_e \rangle$, where $V$ is the set of NFTs as nodes, $T_s, T_e$ are start-time and end-time of observation, $ts : V \rightarrow [T_s, T_e]$ is a function assigning each NFT with a timestamp corresponding to its *primary sell* (or first sell), $E$ is the set of edges such that, for any $v_i, v_j \in V$ *from different collections*, an edge is drawn from $v_i$ to $v_j$ if $v_i$ follows $v_j$, i.e., $ts(v_i) > ts(v_j)$, and $w : E \rightarrow [0, 1]$ is an edge weighing function such that, for any $(v_i, v_j) \in E$, $w(v_i, v_j) = sim(\mathbf{h}_i, \mathbf{h}_j)$, where $sim(\cdot)$ denotes a similarity function for numerical vectors; our default choice is the cosine similarity.

Note that our model definition has two key aspects relating to how NFTs are (i) timestamped and (ii) linked to each other in the graph. As for the first aspect, the availability of the primary-selling timestamps allows us to focus on those NFTs that were traded in the market, thus producing a tangible effect on it. Concerning the second aspect, our rationale for edge drawing and orientation is to capture a notion of *inspiration* triggered by some NFTs for others subsequently appeared in the market. It should be noted that some platforms (e.g., OpenSea) envisage warning mechanisms to alert users about potential "copies" of existing NFTs, however such warnings would be raised at collection level only, thus without actually flagging the involved NFTs. Therefore, identifying issues related to NFT copying, or even plagiarism, remains not trivial to cope with. In this regard, we take a more conservative perspective, as our goal is more generally aimed to detect influential effect of some NFTs towards others by measuring its extent in terms of similarity between the NFT images. Note also that we discard trivial relations of visual inspiration that are likely to occur between pairs of NFTs that belong to the same collection.

COLLECTION GRAPH. Analogously, we define the *NFT Collection visual-inspiration graph* as a directed weighted graph $\mathcal{G}_C = \langle V_C, E_C, ts, w_C, T_s, T_e \rangle$, where $V_C$ is the set of NFT collections as nodes, $E_C$ is the set of edges such that, for any $c_i, c_j \in V_C$, $c_i$ points to (or visually inspired by) $c_j$ (i.e., $(c_i, c_j) \in E_C$) if there exists $n \in c_i$ such that $ts(n) > \min\{t | t = ts(n'), n' \in c_j\}$ (with $ts$ denoting the NFT timestamp assignment function as defined in $\mathcal{G}$), and $w_C : E_C \rightarrow [0, 1]$ assigns each pair of adjacent collection-nodes with a proximity score driven by a principle of NFT nearest-neighbor detection of $c_i$ w.r.t. $c_j$, for any $(c_i, c_j) \in E_C$, according to some pre-defined *collection-linkage criterion*; formally, $w_C(c_i, c_j) = \sigma(\{\max_{n_a \in c_i} sim(\mathbf{h}_a, \mathbf{h}_b), \forall n_b \in c_j \mid ts(n_a) > ts(n_b)\})$, where $\sigma$ denotes a set-function implementing the linkage criterion. To completely specify the above function, we devise three linkage criteria, which correspond to as many ways of capturing different sensitivity levels w.r.t. the NFT inspiration processes across collections:

- $\sigma = \min$, which represents the most tolerant scenario, as it requires that all the involved NFTs reach a certain similarity threshold to hypothesize that the source collection was inspired by the target one;
- $\sigma = \max$, which represents the most eager case, as it is enough to have just one NFT reaching the similarity threshold to hypothesize that the source collection was inspired by the target one;
- $\sigma =$ avg, which represents a balanced case, as it requires that, on average, the pairwise similarity must exceed a given threshold to hypothesize that the source collection was inspired by the target one.

In line with the NFT graph definition, $sim(\cdot)$ would be the cosine similarity between two NFT-collections, ranging between 0 and 1; however, to properly account for the amount of NFTs belonging to an inspired collection $c_i$ that do not contribute to the similarity to another collection $c_j$ (the inspiring one), we introduce a penalization factor of the cosine similarity defined as the sigmoid $1/(1 + \exp(-p_{i \rightarrow j}/np_{i \rightarrow j}))$, where $p_{i \rightarrow j}$ is the number of NFTs in $c_i$ that are regarded as similar to others in $c_j$ (i.e., that are involved in the application of $\sigma$), and $np_{i \rightarrow j}$ is the difference between the total number of NFTs in $c_i$ and $p_{i \rightarrow j}$.

Table 1: Main structural characteristics of the NFT visual-inspiration graphs (left subtable) and similarity-linkage variants of the latest Collection graph (right subtable). Symbol * refers to statistics calculated by discarding edge orientation.

| | NFT-GRAPH | | | COLL-GRAPH (*mid* '21) | | |
|---|---|---|---|---|---|---|
| | *mid* '21 | *mid* '20 | *mid* '19 | max | avg | min |
| #Nodes | 52 198 | 19 011 | 3564 | 488 | 190 | 158 |
| #Edges | 481 873 | 122 046 | 15 501 | 3461 | 612 | 324 |
| Density | 2e-04 | 3e-04 | 0.001 | 0.015 | 0.017 | 0.013 |
| Avg. In-Degree | 9.232 | 6.420 | 4.349 | 7.092 | 3.221 | 2.051 |
| Degree Assortativity | -0.14 | -0.192 | -0.300 | -0.304 | 0.162 | -0.105 |
| %Sources | 8.17 | 11.66 | 17.23 | 43.03 | 43.16 | 53.80 |
| %Sinks | 72.78 | 75.49 | 74.44 | 12.5 | 22.63 | 17.72 |
| Diameter | 17 | 11 | 5 | 9 | 5 | 3 |
| Avg. Path Length | 5.059 | 3.315 | 1.642 | 3.345 | 1.695 | 1.53 |
| Transitivity* | 0.054 | 0.025 | 0.002 | 0.18 | 0.334 | 0.207 |
| Clust. Coeff. * | 0.088 | 0.042 | 0.011 | 0.412 | 0.437 | 0.432 |
| Clust. Coeff. *(full avg)** | 0.062 | 0.029 | 0.007 | 0.357 | 0.336 | 0.315 |
| #SCCs | 52198 | 19011 | 3564 | 452 | 182 | 157 |
| #WCCs | 97 | 36 | 18 | 1 | 8 | 7 |
| #Comm. by *Louvain** | 138 | 69 | 37 | 6 | 14 | 14 |
| Modularity by *Louvain* * | 0.753 | 0.832 | 0.818 | 0.44 | 0.571 | 0.635 |

[8]https://huggingface.co/google/vit-base-patch16-224

Note that the final value of $sim(\cdot)$ still ranges within [0, 1], therefore edges on both the NFT and collection graphs are required to have a similarity score $\geq 0.5$.

## Analysis of the NFT and Collection Networks

In this section, we answer our **RQ2** through a structural analysis of the NFT and Collection networks, respectively, shedding light on their main macroscopic and mesoscopic traits. To this purpose, we focus on *cumulative yearly* observations, so that given our available data, a graph at time $t$ models similarity relations between NFTs, resp. collections, from the beginning (i.e., 2017) up to year $T_e \equiv t$.

**NFT graphs**. Table 1 summarizes statistics on the main structural characteristics of the time-cumulative NFT graphs at mid 2019, mid 2020, and mid 2021. Our choice of starting with a cumulative observation at 2019 has a twofold justification in that more than 80% of the NFTs in our dataset were first sold starting from 2019, while for earlier periods, the similarity relations, and hence the visual inspiration, would be much less significant because of a certain immaturity of the NFT landscape before 2019. Indeed, the rapid growth of the NFT landscape is tangible on the visual-inspiration graphs, with an almost 15x increase in the number of nodes from 2019 to 2021, and an even sharper proliferation of edges in the same observation period (about 31x), hinting at an increasing presence of inspirational mechanisms among the NFT artworks. This is also strengthened by the increased average in-degree over the three-year observation period. Moreover, the negative (directed) degree assortativity, or dis-assortativity, reveals the tendency to draw inspiration from, or to inspire, NFTs having different inspiration degrees.

We also studied the power-law fitting of the in-degrees of the NFT graphs, as reported in the upper rows of Table 2. In this regard, albeit with non-negligible $x_{min}$ values and with a particularly steep curve (see $\alpha$), we spotted hints of likely fitting with the power-law distribution, as also confirmed by the high $p$-values obtained on the *Kolmogorov-Smirnov*'s test (KS-test). These values hold for all the considered timestamps (i.e., from 2019 to 2021) and might shed light on the existence of latent yet perceivable preferential attachment mechanisms on the inspirational process between NFTs.

Interesting remarks arise from the percentages of source and sink nodes, i.e., nodes having no in-links and out-links, respectively. The fraction of sink nodes remains relatively constant and high over the years, denoting that a large part of the NFTs in our graphs are inspiring. The same does not hold for the source nodes, where a progressive decrease in the number of such nodes during the last three years denotes that the remaining majority of NFTs in the network ($\sim$91% of nodes at mid 2021) have inspired other NFTs. This could reasonably be ascribed to the reaching of a sort of *temporary saturation* threshold of NFT artistic development, as most of the visual traits of NFTs could have been used at least once in the market. Such insights are reinforced by the diameter and average path length statistics, which, doubling year by year, reveal how the chain of inspiration was getting longer.

As concerns the triadic closure, both its expressions in terms of transitivity and local clustering coefficient show low values, suggesting that the NFT inspiration (i.e., linkage) process according to visual features could be highly targeted. This is also emphasized by the large and growing number of connected components and communities. In particular, coupled with the observed high modularity, the latter supports the finding on specialized and well-separated inspiration mechanisms among NFTs. In this regard, we notice how the observed communities appear to be induced by specific "visual topics" shared by their corresponding NFTs, such as virtual pets, vehicles, wearable, metaverse lands, gaming cards, and naming services.

Table 2: Outcomes of KS-tests on the power-law fitting of the in-degree distributions of NFT and Collection graphs.

| | graph | $\alpha$ | $x_{min}$ | $p$-value |
|---|---|---|---|---|
| NFT-GRAPH | *mid* 2021 | 4.94 | 168 | 0.81 |
| | *mid* 2020 | 4.11 | 74 | 0.96 |
| | *mid* 2019 | 5.06 | 45 | 0.99 |
| COLLECTION-GRAPH (*mid* 2021) | max | 2.31 | 23 | 0.99 |
| | avg | 2.70 | 12 | 0.99 |
| | min | 2.29 | 7 | 0.99 |

In Figure 1, we provide an illustration of the NFT graph evolving from 2019 to 2021. Looking at the plots, a first remark that stands out concerns the asset outbreak in the transition between years, especially between 2020 and 2021. The "Big Bang"-like visual effect we can notice in the figure gives evidence that inspiration processes have suddenly occurred in the NFT landscape. In particular, the inspiration process have recently involved almost all NFT categories (as evidenced by the colored out-links), eventually coming up in 2021 with *Art*, *Metaverse*, *Games*, and some *Collectibles*, being the most inspiring ones. Therefore, we can reasonably state that the NFT inspiration phenomenon has allowed the market to undergo a rapid expansion.

**Collection graphs**. To analyze the NFT visual-inspiration relations at collection level we focus on the latest time-cumulative models (i.e., mid 2021) by varying the linkage criterion function. A summary of structural statistics is reported in the right-most subtable of Table 1.

As expected, the linkage criteria have a clear impact on the network size, with the average-linkage balancing between the expansion and contraction effects of the max- and min-linkage, respectively. In contrast to what observed for the NFT graphs, the percentage of reciprocated edges is not zero, under all linkage criteria (1.1%, 2.6%, and 0.6% for max, avg, min, respectively), revealing that during different timestamps, cross-collection inspiration events among NFTs of different collections are possible. As previously observed in the NFT graphs, the low density indicates a certain "specialization" of the inspiration process, which is again confirmed by quite a small average in-degree, and conversely, and extremely high number of strongly connected components (almost matching the number of nodes in a graph).

Remarkably, testing the power-law fitting of the in-degree distribution in the Collection graphs, we found clues of preferential attachment, as indicated by the $p$-values resulting from the corresponding KS-tests reported in Table 2. Since the observed goodness-of-fit consistently holds for all link-age criteria, this would suggest that the inspiration process can be pulled from some particularly inspiring collections and creators.

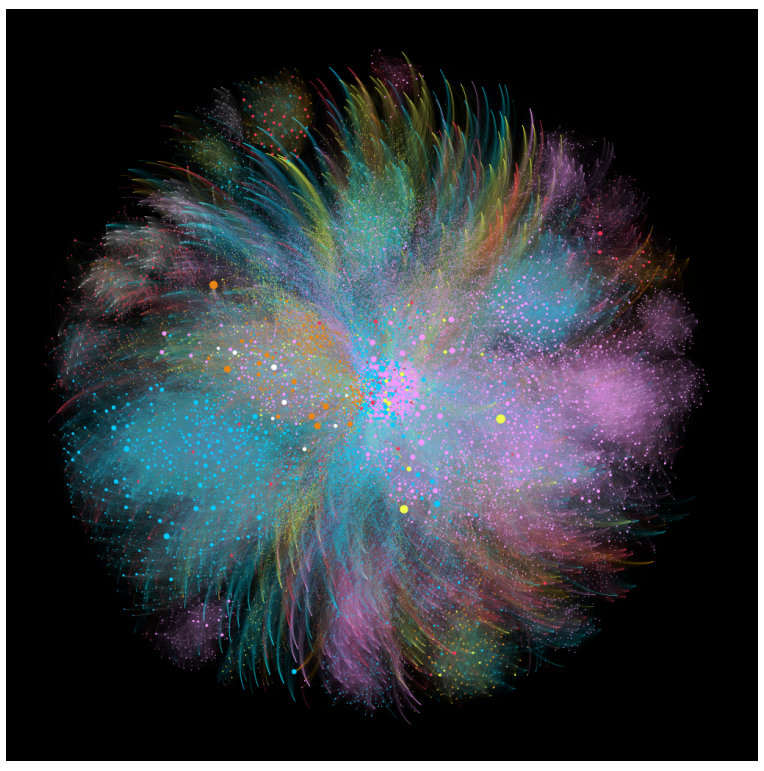
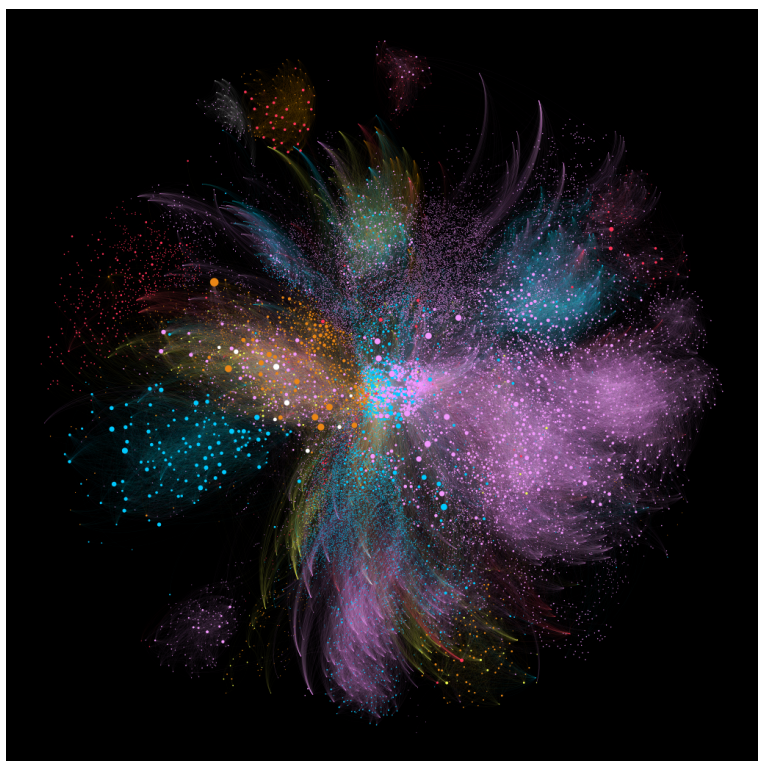
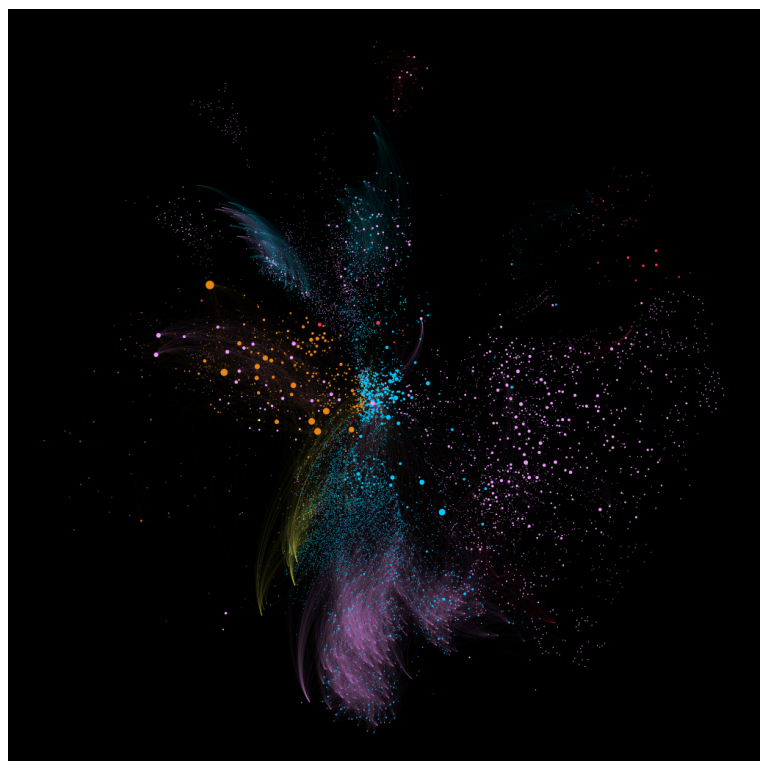

Figure 1: NFT visual-inspiration graphs at 2021 (left), 2020 (center) and 2019 (right). Node colors correspond to categories (pink for Games, yellow for Collectibles, blue for Art, orange for Metaverse, white for Utility, red for Other). Links are colored according to the color of the source nodes. Node size is proportional to the sum of weights on the incoming links to the node; to ease observing the graph evolution, the size of each node is kept fixed to the node's status at 2021. *(Best viewed in color)*

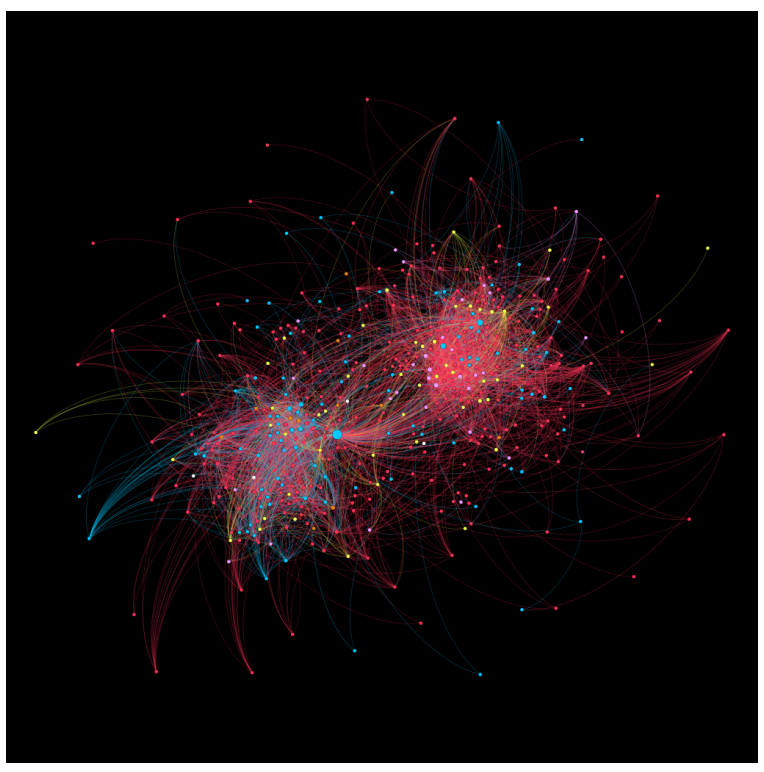
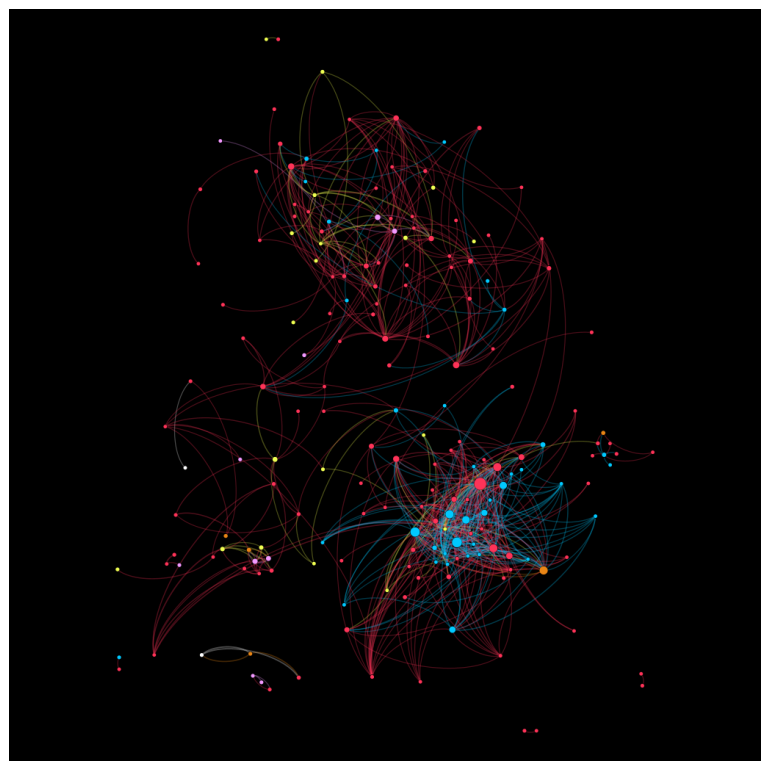
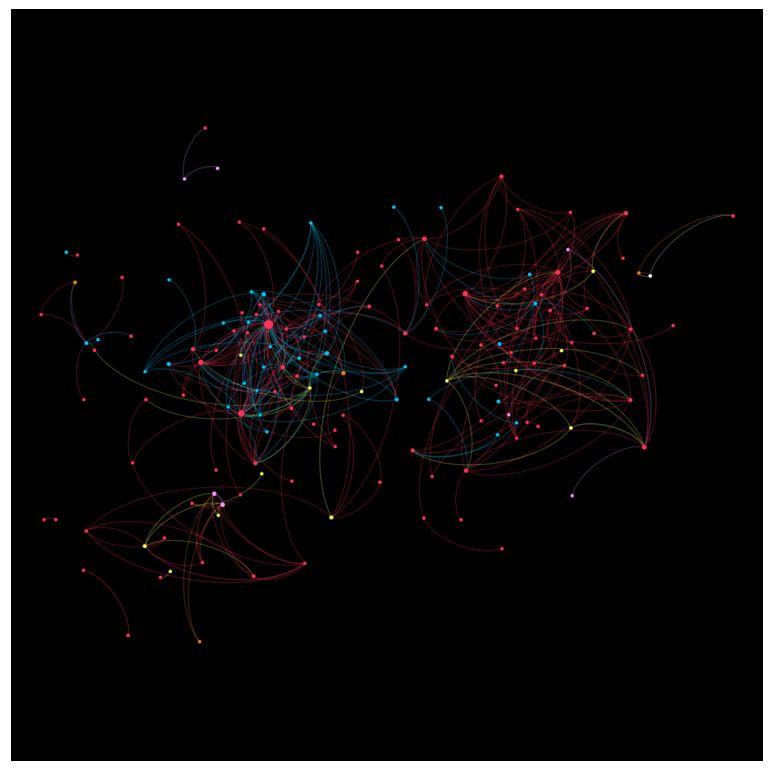

Figure 2: Collection graphs at 2021 by varying linkage criterion: max (left), avg (center), min (right). Node size is proportional to the sum of weights on the incoming links to the node. Color setting is the same as for Figure 1.

The fraction of source nodes, resp. sink nodes follow relatively similar trends under all the linkage strategies, although in an opposite way w.r.t. the NFT graphs: indeed, we spotted that a large fraction of the set of collections actually draw inspiration from others, while the inspiring ones (i.e., those having zero out-degree) represent a minority of the total collections involved in inspiration processes.

The choice of linkage criterion also impacts on the degree assortativity. According to the graph definition, the disassortativity corresponding to the *max* case indicates that nodes having a high out-degree (i.e., collections inspiring to many others) tend to get also inspired by nodes with small in-degree (i.e., collections that inspire little); this is clearly in accord with the high sensitivity of the aforementioned linkage criterion, which is, therefore, able to spot even the subtlest semblances of inspiration. On the contrary, the disassortativity corresponding to the *min* case allows us to reasonably state that collections drawing little inspiration from others (i.e., small out-degree) do this by targeting the most inspiring collections (i.e., high in-degree); again, this fits the behavior due to the linkage criterion, which is here the most cautious in terms of similarity recognition. The two above contrasting behaviors are finally balanced by the *average*-linkage criterion, which leads to a positive degree correlation, which means that, on average, collections that are very inspired do it from those that are very inspiring, thus hinting again at a preferential attachment mechanism, consistently with the previously discussed finding (cf. Table 2).

In addition, the values of diameter and average path length, regardless of the linkage criterion, further shed light on the close-knit process of inspiration among collections. The linkage of collections hence appears to shape a tight network system, albeit sectorial, as indicated by triadic closure and modularity-based statistics. In light of such remarks and the previous findings about the NFT graphs, we again report a notable and even more evident matching between community structures and specific visual traits.

Table 3: Market-related statistics for the inspiring and inspired NFTs computed from the 2021 NFT visual-inspiration graph.

| Financial Indicator | Inspiring NFTs | Inspired NFTs | Inspiring/ Inspired |
|---|---|---|---|
| average volume | \$231531.69 | \$146192.15 | 1.584 |
| average #transactions | 151.92 | 100.00 | 1.519 |
| average price | \$692.91 | \$899.09 | 0.771 |
| maximum price | \$6661.95 | \$4605.24 | 1.447 |
| minimum price | \$102.24 | \$318.89 | 0.321 |
| st. dev. price | \$977.22 | \$725.69 | 1.347 |

In Figure 2, we show the effect of the linkage criteria on the 2021's Collection graph. At first glance, it appears that the $\min$ linkage criterion tends to trigger scattered links among collections, thus leaving apart potential cases of mild or fairly moderate inspirational scenarios. Conversely, the eagerness of the $\max$ linkage criterion yields a proliferation of similarity links, forming two close-knit and dense "clouds" of visual inspiration. Besides, the avg linkage approach balances between the aforementioned strategies, overcoming the extreme sparseness of the $\min$ case, thus effectively being able to capture actual similarities, yet without degenerating into potential noise of the $\max$ case. Moreover, regardless of the chosen linkage criterion, some shared traits emerge from Figure 2: *Art* and *Collectibles* appear to be the most inspiring categories, whereas *Other* results to be the one corresponding to the most inspired NFTs (red outlinks), by also being the most generic ones — thus including any potential asset. Interestingly, at mid-2021, *Games* and *Metaverse* appear to have carved out a portion of the market away from other visual features, or with an original redefinition of the same, as their presence is almost latent in the plots.

## Market-based Characterization of the NFT Visual Inspiration Phenomenon

Our previous analysis has shed light on the structural traits of the visual-inspiration networks built on the NFTs and their collections. Here we want to provide a simple characterization of the dichotomy between *inspiring* and *inspired* NFTs in terms of main aggregated financial indicators (**RQ3**).

Based on the latest update of the NFT visual-inspiration graph (i.e., 2021), Table 3 reports, for various market-related statistics, the ratio between inspiring NFTs (target nodes in a link) and inspired NFTs (source nodes in a link).

We first notice that inspiring NFTs are able to achieve more transactions and higher volumes[9] than inspired ones, with peaks of $+58\%$ and $+52\%$, respectively. Surprisingly, we find lower average, resp. minimum, prices for inspiring NFTs, since inspired NFTs obtain a $+23\%$, resp. $+68\%$ advantage over the former. We shed light on such a counter-intuitive trait by looking at the ratio of standard deviations, which shows a $+35\%$ for inspiring NFTs. As a result, we can state that the inspiring NFTs face an initial struggle in the market with lower minimum prices due to the novelty of the proposed visual traits, with a more evident price instability, whereas inspired ones tend to enter a market "ready" to appreciate the proposed features, thus starting from higher minimum prices. Nonetheless, we also spotted that the market inherently regulates such a phenomenon, with inspiring NFTs able to achieve higher maximum prices w.r.t. inspired ones ($+45\%$), thus conferring a sort of competitive advantage to the former.

[9]In the NFT market, volume of an NFT is meant as the total amount sold for all transactions involving that NFT.

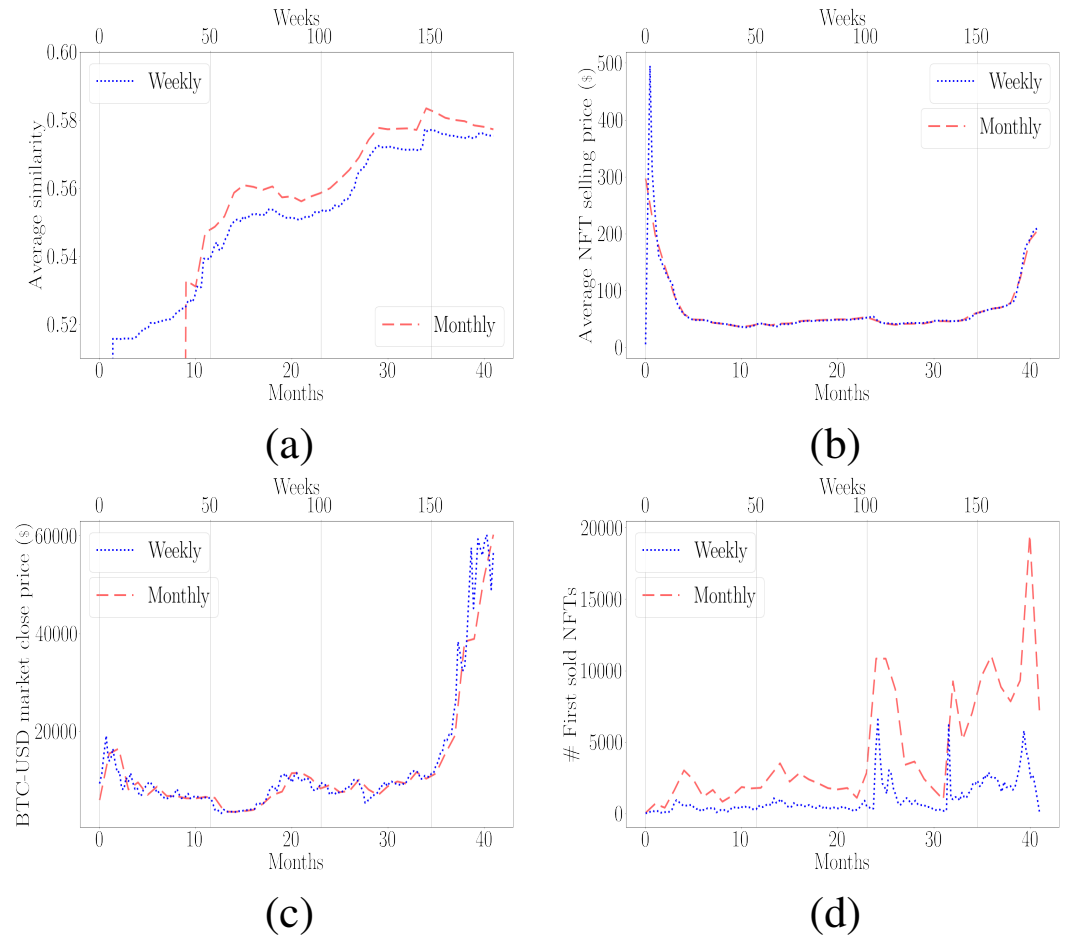

Figure 3: Average NFT similarities over time (a), average NFT selling prices (b), Bitcoin market closing prices (c) and number of first-sold NFTs (d).

## Crypto Influence Dynamics

Answering our **RQ4** requires to analyze the temporal relation between trends of market-related indicators and trends of NFT-image similarities. To this purpose, we resort to the Time Lagged Cross Correlation (TLCC) technique which, given two time series $s$ and $s'$ of the same length, determines how well $s'$ is related to past lags of $s$, thus identifying lags of $s$ that might be predictors of $s'$. Formally, the TLCC of $s, s'$ is the set of sample Pearson-correlations between $s_{t+\ell}$ and $s'_t$, for all $\ell \in [-T..T]$ $(T > 0)$, at any given time-step $t$ within a chosen time period $T$. The lag refers to how far the series are offset, and its sign indicates which series is shifted in time, therefore $\ell < 0$ corresponds to a correlation between $s$ at a time before $t$ and $s'$ at time $t$, i.e., $s$ leads $s'$, whereas $\ell > 0$ corresponds to a correlation between $s$ at a time after $t$ and $s'$ at time $t$, i.e., $s$ lags $s'$.

Within this view, we analyze different types of TLCC to assess whether a leading-lagging relation exists between "artistic" and "market-related" indicators. In particular, using either a *monthly* or a *weekly* sampling — i.e., time-steps correspond to either months or weeks — we modeled five types of time series over the full period of observation (i.e., $T_e - T_s$), measuring for each time-step $t$: (i) the average of the pairwise similarities (including both within- and across-*category* pairs) of NFTs from the NFT graph observed up to $t$ (i.e., $T_e = t$), (ii) the average of the NFTs' mean selling prices up to $t$, (iii) the closing price of the Bitcoin market at $t$, (iv) the number of first-sold NFTs at $t$, and (v) the number of collections containing at least one first-sold NFT, at $t$.

**NFT and market-related time series.** Before moving to the analysis of the TLCCs, let us first provide insights into the shapes of the aforementioned time series. (For the sake of brevity, we discard a discussion about the series of collections containing first-sold NFTs; however, main remarks for the latter hold similarly to those for the plots of first-sold NFTs). As shown in Figure 3-(a), the average pairwise similarity series exhibit a similar and generally increasing trend, although interesting differences at their starting point arise depending on the time sampling of the evolving NFT visual-inspiration graph. In fact, the weekly series shows non-zero similarities since the first week of 2018, whereas the monthly one "delays" the raising of visual inspiration events to August 2018. For both series, we also notice an abrupt growth in similarities around the 10th month, before the trend becoming slower or even stopping around the 30th month. As a further remark (not drawn from Figure 3-(a)), although the average growth rate turns out to be more evident for the within-category similarity, the trend of the average similarities follows that of the across-category similarities.

Looking at the average NFT-selling price series, shown in Figure 3-(b), we notice almost identical shapes from the monthly and weekly samplings. Interestingly, the peaks at the beginning of the series correspond to the dawn of the NFT advent. This should be ascribed to the scarcity of assets yet the novelty of this technology, which dictated those high entry-prices. Then, a large time window follows as characterized by low prices, which can be explained by the strong proliferation of NFTs in the market. Eventually, a rising trend is observed during 2021, when the market achieved its maturity, and the *fomo* (i.e., fear-of-missing-out) begins to spread over more people. It should also be noted how the two discussed plots are inter-related, hinting at self-regulation of the market, where similarities and average prices appear to be (inversely) affected by each other.

Considering Figure 3-(c), the initial upward trend is rapidly stopped by the bear market where a minimum of \$3500 is reached at the beginning of January 2019 (14th month, resp. 58th week). A new trend inversion then begins, thus increasing the BTC price to around \$9000 for a time span that goes from the 20th month to the 35th month, then again the bull-run drives the BTC price upward reaching about \$60000 at the end of the observation interval.

Finally, the upward trend characterizing both series in Figure 3-(d), appears to be more prominent for the monthly sampling. Two distinct peaks are noticed at around the 40th month, resp. 100th week, corresponding with the trend inversion of the BTC bear/bull-run and the start of the bull-run of 2020/2021.

**Time-lagged cross correlations.** We are now ready to analyze the TLCCs between the time series previously discussed, where we specify $T$ equal to one year. Figure 4 shows the TLCC between average NFT selling prices and average NFT similarities, and the TLCC between average NFT-selling prices and Bitcoin market closing prices, where the latter quantity in either comparison corresponds to $s'$ in the definition of TLCC reported at the beginning of this section.

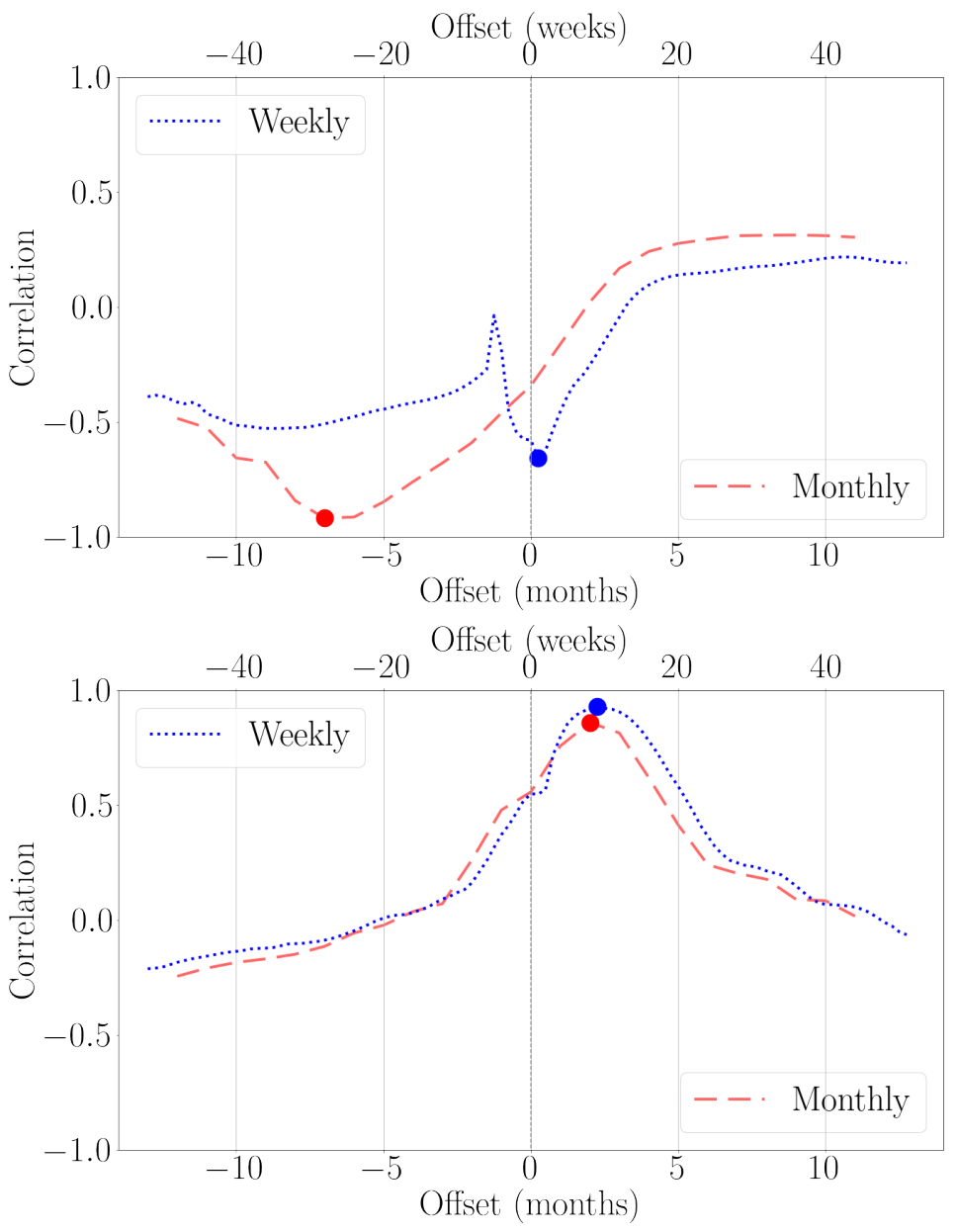

Figure 4: Time lagged cross-correlation of average NFT-selling prices vs. average NFT similarities (top), and of average NFT-selling prices vs Bitcoin market closing prices (bottom). In each plot, the bottom x-axis corresponds to monthly samples (red dashed line), whereas the top x-axis corresponds to weekly samples (blue dotted line).

Considering the first comparison (top correlogram in Figure 4), according to a monthly sampling we find a strong negative correlation ($-0.918$) between the average NFT selling prices and the average NFT similarities, with a negative offset ($t = -7$ month). This means that in the mid term, selling prices lead the NFT inspiration process in an inverse way, i.e., high prices reduce similarities, and hence lower the tendency of reusing visual features (i.e., temporary saturation of artistic development), and vice versa. Conversely, the weekly perspective sheds light on a short-term strategy with peak negative correlation ($-0.658$) at $t = 1$ week, thus indicating that the selling prices lag the NFT similarities, still in an inverse way. Remarkably, this finding unveils that the temporary saturation of artistic development can impact on asset performances by reducing the NFT prices. Besides, as the leading quantity changes depending on the sample resolution (i.e., weekly or monthly), we can spot different latent strategies that would be triggered by the market in response to the variation of the aforementioned quantities.

The bottom correlogram of Figure 4 shows the TLCC between the average NFT selling prices and the closing prices of the crypto markets, represented via the BTC-USD trends obtained through the Yahoo! Finance APIs.[10] Looking at the plots, two main remarks stand out: (i) BTC-USD is the leading quantity in both medium- and short-term scenarios, with peaks of correlation up to 0.93 and 0.898 for the monthly and weekly perspective, respectively; and (ii) there exists a plateau of correlation according to both weekly sampling (from the 5th to the 17th week) and monthly sampling (from the 2nd to the 5th month). The coherence of such observations across different samplings and the corresponding strong correlation allow us to confirm that BTC is the main influence factor in determining the NFT selling performances in line with other studies (Pinto-Gutiérrez et al. 2022; Ante 2022; Anselmi and Petrella 2023).

[10] Available at https://finance.yahoo.com/

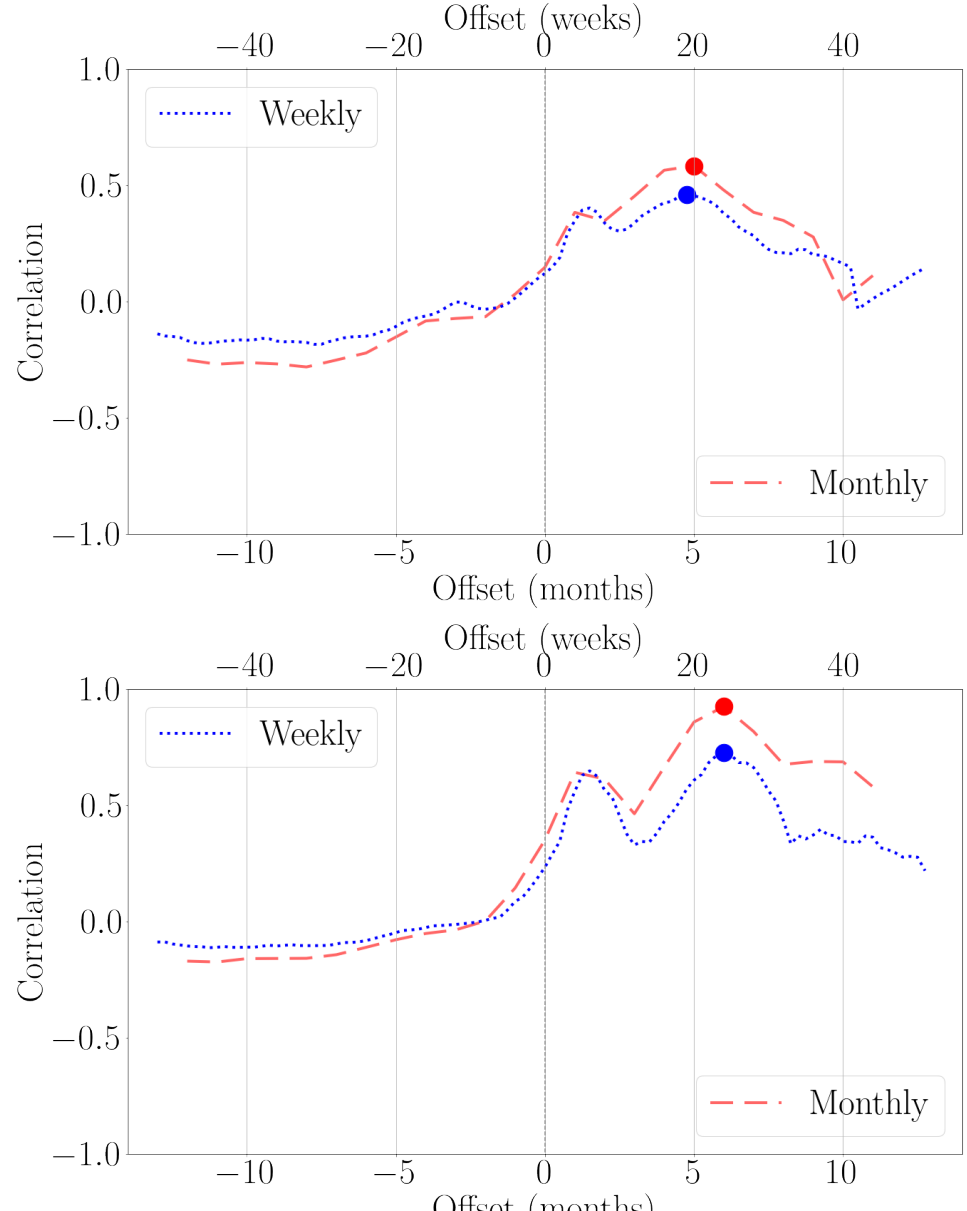

Figure 5: Time lagged cross-correlation of average NFT-selling prices vs. number of first-sold NFTs (top), and of average NFT-selling prices vs. number of collections containing at least one first-sold NFT. In each plot, the bottom x-axis corresponds to monthly samples (red dashed line), whereas the top x-axis corresponds to weekly samples (blue dotted line).

Furthermore, in Figure 5, we show the TLCC between prices and counts of newly created NFTs/collections on both weekly and monthly offsets. Besides spotting an almost equivalent offset between months or weeks for either NFTs and collections, we find strong evidence that their birth impacts on the market. More precisely, collections appear to be a better financial predictor due to the greater volume they generate, and such an indicator is stronger when we consider a monthly perspective, with peaks of correlation up to 0.92 and 0.58 for collections and NFTs, respectively.

## Explainability Aspects of the NFT Visual Learning Model

In this section we aim at answering our **RQ5**, which allows for gaining qualitative insights into the ability of our proposed approach to shape similarities between NFTs, thus providing an explanation underlying a relation of visual inspiration between any two NFT images.

To this purpose, we resort to the SHAP (SHapley Additive exPlanation) framework (Lundberg and Lee 2017),[11] which leverages the game-theoretic Shapley value to locally explain the outcome of a machine/deep learning model, like our NFT visual-inspiration learner, whereby input features are represented through players of the game. SHAP is a model-agnostic technique, since it just requires class probabilities generated by the learner and, by selectively perturbing the learner's input features, it calculates the contribution that each of the features gets w.r.t. the output probabilities. Details on SHAP are given in ***Appendix***.

In Figure 6, we report three exemplary cases, sampled by linked NFTs in our 2021 NFT-graph, and ordered by left to right with increasing level of difficulty for our NFT visual-inspiration learner. Let's begin with two NFTs representing the Van Gogh's *A Pair of Shoes* painting, shown in Figure 6-(a). As expected, since the two images only differ in terms of color saturation and contrast, our model was able to identify a strong similarity between them. This similarity is well explained by SHAP as almost all portions of the shoes are detected as relevant features (red cells).

The case depicted in Figure 6-(b) is an example of "extreme" inspiration captured by our NFT visual-inspiration learner, since the two NFTs display a common concept, i.e., the handle pickaxe, by an almost identical drawing design, even though with different color choices for their heads. Indeed, the SHAP model correctly detects the "handle" portions and the "Vs" symbols of the images as relevant features for the similarity between the two NFTs (red cells), whereas the "heads" are detected as irrelevant features (blue cells). Also, the textual components of the two NFTs ("Gold" and "Kryptonite") are correctly discarded by SHAP in explaining their similarity. It should be emphasized that, like in other cases, the two NFTs are from different collections.

Figure 6-(c) shows two apparently different NFTs, which nonetheless share a common visual concept concerning a warrior figure. The inspiration exerted by the rightmost NFT to the leftmost NFT is well captured, as indicated by the higher Shapley values (i.e., red cells) corresponding to the upper portion of the warrior on the left figure and the torso of the second one. Furthermore, the model is able to distinguish the different weapons (i.e., a katana on the left, and a sword on the right), as indicated by the blue cells located upon their respective locations. Therefore, not only our NFT visual-inspiration learner effectively captures similarity (0.74) underlying the concept of "warrior", but does so by ignoring pointless features (e.g., text over the left image) and, most importantly, regardless of the observation perspective (front with zoom inset on the left, or in profile on the right).

[11] Code available at https://github.com/slundberg/shap

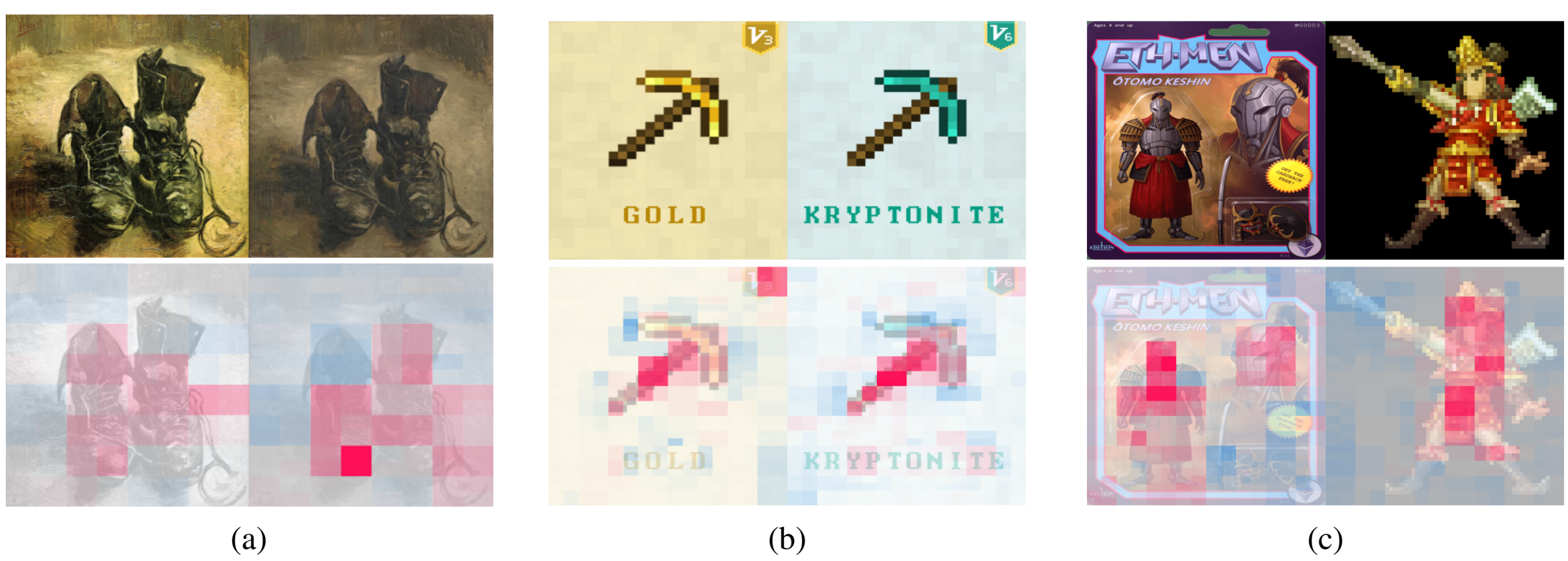

Figure 6: Use cases for explanation of our PVM learner for NFT images. Each case consists of two images placed side by side that correspond to NFTs linked in our inspiration-networks, and hence from different collections (top), along with their SHAP explanations (bottom) in the form of heatmap layers. Blue, resp. red, cells denote negative, resp. positive, impact of features on the model's similarity prediction.

## Related Work

The scientific community is paying increasing attention to the decentralized paradigm and the advent of Web3, with the goal of seizing new opportunities and managing emerging challenges (Cao 2022; Chen et al. 2022).

Focusing on NFTs, (Wang et al. 2021) were among the first to provide an in-depth technical perspective of protocols, standards, and desired properties for the NFT technology. Similarly, in their seminal work, (Nadini et al. 2021) delved into more than 6M transactions concerning the trade of about 5M NFTs between 2017 and 2021 to shed light on the main patterns that characterize NFTs. They revealed the tendency of traders to specialize w.r.t. some specific collections, the visual-feature homogeneity of collections, and how visual aspects might improve the price predictability solely based on transactions.

Other studies have attempted to identify useful predictors for NFT financial performance, leveraging either intrinsic or extrinsic features. As for the former, (Mekacher et al. 2022) explored about 4M transactions, occurred between 2018 and 2022 and involving 1.4M NFTs from more than 400 collections to analyze how visual attribute rarity shapes financial performances. The heterogeneous distribution of rarity across most collections was found to affect selling prices and asset's stability. Moreover, utility-based features (e.g., attack and defence card scores) turned out to improve rarity-based predictions, as found for the *Axies* play-to-earn gaming NFTs (Ho et al. 2022).

As concerns external influence factors, although cryptos might provide some hints at the pricing of NFTs (Dowling 2022; Pinto-Gutiérrez et al. 2022; Ante 2022; Apostu et al. 2022; Anselmi and Petrella 2023), social media seem to be a tangible influence factor for NFT financial performances. In this regard, (Kapoor et al. 2022) linked OpenSea and Twitter data to leverage the linkage between crypto assets and social phenomena for an NFT pricing prediction task, whose evaluations revealed that accounting for social information allows achieving a boost up to 6% in terms of accuracy w.r.t. models exploiting just NFT-related features. Recently, we addressed the challenge of financial performance prediction through the definition of a financial-agnostic and multimodal deep learning framework that exploits the hypothesis that NFT images and descriptions might serve as proxies for their selling prices (Costa, La Cava, and Tagarelli 2023). The network-based approach in (Vasan, Janosov, and Barabási 2022) focuses on the Foundation platform through more than 48k NFTs involving about 15k artists, showing that the linkage between artists and collectors is a determinant for the development of a tight and enduring network of investments. In particular, the "arrival time" in the platforms for both artists and investors shapes their earnings and spending, respectively. Also, although NFT prices are subject to fluctuations, they tend to be within a stable range for each creator, thus determining her/his reputation. The relevance of the above aspects is further confirmed by (Colavizza 2022). By analyzing the seller-buyer networks of about 40k sales, the authors unveiled that the NFT market is driven by a small set of sellers and (an even smaller) set of buyers. Besides, they reported that preferential buyer-seller ties characterize the growth of the market, and more interestingly, ties persist even during dips, thus becoming a footprint of the NFT landscape growth.

## Conclusions

We presented the first work leveraging visual features learned from NFT images and used to build a network model for capturing and analyzing visual inspiration relations between NFTs.

By answering a number of research questions relating to the visual inspiration phenomenon, our results have shown that (i) the inspirational mechanisms underlying the artistic development of NFTs are pervasive and they have progressively led to a temporary saturation of the visual feature space, thus affecting originality of NFTs; (ii) the dichotomy between inspiring and inspired NFTs has effect on related financial indicators; and (iii) there exist latent and enduring mechanisms of self-regulation between markets and inspiration waves.

Several directions can be outlined as future research. One aspect that is part of our ongoing work is the development of a visual learning model specific for the NFT domain, by re-training or further training a PVM like our used ViT, or alternative models, on NFT image data collections. A related opportunity comes from optimizing an NFT visual model to downstream tasks, such as classification and object recognition, possibly in multi-modal learning scenarios.

We would also like to point out that our definition of NFT visual inspiration graph models can straightforwardly be generalized to other datasets of NFTs, which might also include more information than the one used in this study, such as the NFT minting timestamps.

We expect our study on NFT visual inspiration can pave the way for understanding important mechanisms arising during the evolution of Web3.

**Impact and Ethical Considerations**. The aim of this study is to leverage the concept of visual inspiration to shed light on the visual similarity among published NFTs, and on its impact on the NFT market dynamics. Nonetheless, our findings concerning visual inspiration among NFTs should be taken with caution if one would like to use them for addressing critical tasks, such as copyright infringements or others relating malicious market dynamics.

Moreover, we chose not to reference any specific artwork, collection or creator throughout the manuscript to avoid exposing specific creators to accusations of plagiarism or untrustworthiness.

As a final general remark, any finding that might be drawn from this work should be intended to support decision-making during exploring (possibly, trading) NFTs, and not to replace the human specialists.

**Conflict of interest**. The authors declare that they have no conflict of interest and no affiliations with or involvement in any organization or entity with any financial interest or non-financial interest in the subject matter discussed in this manuscript.

# Appendix

## Details on SHAP

SHAP is used to explain the outcomes of a model designed for the following classification task: Given two PVMs with shared weights, and an input pair of tokenized images $(\mathcal{T}_i, \mathcal{T}_j)$, the goal is to predict the probability that the two images are similar or not similar. The class probabilities are defined as $\langle sim(\mathbf{h}_i, \mathbf{h}_j), 1 - sim(\mathbf{h}_i, \mathbf{h}_j)\rangle$, where $sim(\mathbf{h}_i, \mathbf{h}_j)$ is the cosine similarity between $\mathbf{h}_i$ and $\mathbf{h}_j$, i.e., the embedding for image $i$ and $j$, respectively (cf. Sect. *Data Extraction and Network Modeling*).

Shapely regression values are represented as a linear model over importance scores assigned to the features (Lundberg and Lee 2017). In our setting, features correspond to groups of pixels extracted from each input pair of images. These raw features are then corrupted through blurring functions which are aimed at creating two coalitions of images for each given feature: a coalition where a particular feature is present and one where it is corrupted. More precisely, for each feature $f \in F$ (with $F$ indicating the space of features), the Shapley value $\Phi_f$ for $f$ can in principle be computed by estimating the difference between the prediction of a model $\mathcal{M}$ (i.e., pair of PVMs) when the feature is used and the prediction of the model without that feature:

$$\Phi_f = \sum_{S \subseteq F \setminus \{f\}} \frac{|S|(|F| - |S| - 1)}{|F|} [\mathcal{M}_{S\cup\{f\}}(x_{S\cup\{f\}}) - \mathcal{M}_S(x_S)]$$

where $x$ denotes a feature representation of the pair of images which depends on each of the *coalitions* $S$ selected from the feature space $F$. To efficiently computing the explanations, Shapely values are estimated by sampling approximations, thus allowing to avoid computing $2^{|F|}$ possible coalitions and retraining the model $\mathcal{M}$.

Before computing explanations, we resized each image to $512 \times 512$ pixels. Upon this, we configured SHAP to compute $K = 10\,000$ samples for each explanation, where a Gaussian kernel of size $64 \times 64$ is used to mask features by generating a Gaussian blurring.